\documentclass[prd,showpacs,showkeys,nofootinbib,12pt
]{revtex4}
\usepackage{epsfig}

\def\a{\alpha}

\def\vta{\vartheta}

\begin{document}

\title{Dimensions and Units in Electrodynamics}

\author{Friedrich W.\ Hehl$^{1,2}$\email{hehl@thp.uni-koeln.de}, Yuri
  N.\ Obukhov$^{1,3}$\email{yo@thp.uni-koeln.de}}

\affiliation{$^1$ Institut f\"ur Theoretische Physik, Universit\"at zu
  K\"oln, 50923 K\"oln, Germany   \\
  $^2$ Department of Physics and Astronomy, University of
  Missouri-Columbia, Columbia, MO 65211, USA\\
$^3$ Department of Theoretical Physics,
  Moscow State University, 117234 Moscow, Russia}

\date{06 July 2004, {\em file okun10.tex}}

\begin{abstract}
  We sketch the foundations of classical electrodynamics, in
  particular the transition that took place when Einstein, in 1915,
  succeeded to formulate general relativity. In 1916 Einstein
  demonstrated that, with a choice of suitable variables for the
  electromagnetic field, it is possible to put Maxwell's equation into
  a form that is covariant under general coordinate transformations.
  This unfolded, by basic contributions of Kottler, Cartan, van
  Dantzig, Schouten \& Dorgelo, Toupin \& Truesdell, and Post, to what
  one may call {\em premetric classical electrodynamics.} This
  framework will be described shortly. An analysis is given of the
  physical dimensions involved in electrodynamics and subsequently the
  question of units addressed. It will be pointed out that these
  results are untouched by the generalization of classical to quantum
  electrodynamics (QED). We compare critically our results with those
  of {\sl L.B.\ Okun\/} which he had presented at a recent conference.
\end{abstract}



\maketitle

\newpage
\section{Introduction}

When Maxwell, in the 1860s, formulated the field equations of
electrodynamics,\footnote{For detailed historical accounts and for the
  corresponding references, see Darrigol \cite{Darrigol} and Whittaker
  \cite{Whittaker}.} he used the fields $E,H,D,B$. Sometimes he would
substitute $B$ by $\mu H$. Subsequently, around 1900, Lorentz in his
electron theory reformulated Maxwell's equations and eliminated (in a
paper of 1904, see \cite{Lorentz1904}) $E$ and $B$ by putting, in
vacuum, $E=D$ and $B=H$.

This theory, often called Maxwell-Lorentz theory, had difficulties,
like Maxwell's original version, in explaining the aberration of
light, which had been discovered by Bradley in 1729. Already Hertz,
R\"ontgen, and others tried to formulate electrodynamics for moving
matter. The outcome of the Michelson-Morley experiment, which had been
set up since Michelson knew about the aberration problem, enforced a
reformulation of the Maxwell-Lorentz equations, or rather their
transformation behavior by extending the Galilean transformations of
classical mechanics.

Finally, Einstein in 1905 resolved the aberration problem, inter alia,
in his paper ``On the electrodynamics of moving bodies.'' Minkowski
(1908) put this {\em special relativity theory\/} (SR) in its final
form by introducing the concept of a flat 4-dimensional
pseudo-Euclidean spacetime continuum, the Minkowski space(time).  

This paper is dedicated to our colleague and friend {\sl Alberto A.
  Garc\'{\i}a} (CINVESTAV, Mexico City) on the occasion of his 60th
  birthday.

\section{Poincar\'e covariant electrodynamics}

The group of motion in Minkowski spacetime is the Poincar\'e group
(also known as inhomogeneous Lorentz group) that is a semidirect
product of the translation group and the (homogeneous) Lorentz group.
The Maxwell-Lorentz theory was then put in a Poincar\'e covariant
form, see, for example, Einstein's presentation in \cite{meaning}:
\begin{equation}\label{meaningSR}
  \frac{\partial\phi_{\mu\nu}}{\partial x_\nu}={\cal J}_\mu\,,\qquad
  \frac{\partial\phi_{\mu\nu}}{\partial x_\sigma}+
  \frac{\partial\phi_{\nu\sigma}}{\partial x_\mu}+
  \frac{\partial\phi_{\sigma\mu}}{\partial x_\nu}=0\,.
\end{equation}
Einstein used a Euclidean metric with an imaginary time coordinate.
For this reason all indices are covariant ones.

Later, in the 1930s and the 1940s, a quantized version of
electrodynamics with the Dirac electron as a source was developed on
the basis of (\ref{meaningSR}) and the Dirac equation. Quantum
electrodynamics (QED), as the newly emerging theory was called, is a
theory in the framework of SR and the Poincar\'e group. The
overwhelming success of QED made it look as if (\ref{meaningSR}) were
the final answer to an appropriate formulation of the field equations
of classical electrodynamics.

However, such a view neglects the impact Einstein's gravitational
theory, the general theory of relativity (GR), had on the
understanding of the structure of Maxwell's equations.

\section{Electrodynamics after the advent of general relativity}

Immediately after Einstein's fundamental 1915 paper on GR and even
before his big survey paper on GR would appear, Einstein
\cite{Einstein1916} observed that Maxwell's equations can be put in a
general covariant form by picking suitable field variables. This
meant, unnoticed even today, more than 80 years later, by most
elementary particle physicists, the reanimation of the $D$ and the $B$
of Maxwell (or, with Lorentz's choice, of $E$ and $B$).

Einstein \cite{Einstein1916} wrote Maxwell's equations
as\footnote{Einstein used subscripts for denoting the coordinates $x$,
  i.e., $x_\tau$ etc.  Moreover, we dropped twice the summation symbols
  $\Sigma$.}
\begin{equation}\label{akademie}
  \frac{ \partial F_{\rho\sigma}}{\partial x^\tau}+ \frac{ \partial
    F_{\sigma\tau}}{\partial x^\rho}+ \frac{ \partial
    F_{\tau\rho}}{\partial x^\sigma}=0\,,\quad {\cal
    F}^{\mu\nu}=\sqrt{-g}g^{\mu\alpha}
  g^{\nu\beta}F_{\alpha\beta}\,,\quad \frac{\partial{\cal
      F}^{\mu\nu}}{\partial x^\nu}={\cal J}^\mu\,,
\end{equation}
In his ``Meaning of Relativity'' \cite{meaning}, in the part on GR ---
often overlooked by aficionados of the two Poincar\'e covariant
equations in (\ref{meaningSR}) --- he picked the letter $\phi$ for the
field strength apparently in order to stress the different nature of
${\cal F}$ and $\phi$:
\begin{equation}\label{meaningGR}
  \frac{ \partial \phi_{\mu\nu}}{\partial x^\rho}+ \frac{ \partial
    \phi_{\nu\rho}}{\partial x^\mu}+ \frac{ \partial
    \phi_{\rho\mu}}{\partial x^\nu}=0\,,\quad \frac{ \partial {\cal
      F}^{\mu\nu}}{\partial x^\nu}={\cal J}^\mu\,,\quad {\cal
    F}^{\mu\nu}=\sqrt{-g}g^{\mu\sigma} g^{\nu\tau}\phi_{\sigma\tau}\,.
\end{equation}

On the foundations of GR, besides the equivalence principle, there
lays the principle of general covariance. And the Maxwell equations
(\ref{meaningGR})$_1$ and (\ref{meaningGR})$_2$ are generally
covariant {\em and\/} metric independent. Since in GR the metric $g$ is
recognized as gravitational potential, it is quite fitting that the
fundamental field equations of electromagnetism do {\em not\/} contain
the gravitational potential.  

The gravitational potential only enters equation
(\ref{meaningGR})$_3$. We call this equation the spacetime relation
--- it is the ``constitutive law'' of the vacuum. Needless to say that
having understood that there exists a way to formulate Maxwell's
equations in a generally covariant and metric-independent manner,
going back to (\ref{meaningSR}) would appear --- since 1916, in fact
--- to be an anachronism.

The poor man's way to adapt the Poincar\'e covariant equations in
(\ref{meaningSR}) to the Riemannian spacetime of GR is to substitute
the partial derivatives $\partial$ by covariant ones $\nabla$. Then,
after some algebra, one also ends up at equations equivalent to
(\ref{meaningGR}). However, one wouldn't understand why the Maxwell
equations are generally covariant at all. Moreover, the excitations
$\cal D$ and $\cal H$ wouldn't show up. Besides the field strengths
$E$ and $B$, defined via the Lorentz force, the excitations $\cal D$
and $\cal H$ can be directly {\em measured\/} also in vacuum by the
Maxwellian double plates and the Gauss method, respectively. In other
words, $\cal D$ and $\cal H$, similar to $E$ and $B$, do have an own
operational interpretation, see also \cite{PhysicsWorld,gentle}.

Following Feynman \cite{Feynman}, we call $F=(E,B)$ the
electromagnetic {\em field strength\/} and, following Mie \cite{Mie}
and Sommerfeld \cite{Sommerfeld}, $H=({\cal H},{\cal D})$ the
electromagnetic {\em excitation}.  In exterior calculus, switching now
to our conventions \cite{book}, see in this context also the new books
of Lindell \cite{Lindell} and Russer \cite{Russer}, the 4-dimensional
Maxwell equations read
\begin{equation}\label{maxwell}
dH=J\,,\qquad dF=0\,,
\end{equation}
with the decompositions $H=H_{ik}\,dx ^i\wedge dx ^k/2$, $F=F_{ik}\, dx
^i \wedge dx ^k/2$, and $J = J_{ik\ell}\, dx ^i\wedge dx ^k \wedge dx
^\ell/6$, where $dx ^i$ is a natural (or holonomic) coframe, a basis
of the cotangent space. In tensor calculus we have
\begin{equation}\label{maxtensor}
  \partial_k\check{H}^{ik}=\check{J}^i\,,\qquad
  \partial_{[i}F_{k\ell]}=0\,,
\end{equation}
with $\check{H}^{ik}=\epsilon^{ik\ell m}\,H_{\ell m}/2$ and $\check{
  J}^i =\epsilon^{ik\ell m}\, J_{k\ell m}/6$. Here $\epsilon^{ik\ell
  m}$ is the generally covariant Levi-Civita symbol with values $\pm
1,0$. Of course, (\ref{maxwell}) and (\ref{maxtensor}) are just
alternative versions of (\ref{meaningGR})$_1$, (\ref{meaningGR})$_2$.

On the surface of a neutron star, for example, where we might have
strong magnetic fields of some $10^{10}$ tesla and a huge curvature of
spacetime --- we are, after all, not too far outside the Schwarzschild
radius of the neutron star of some 3 km --- the Maxwell equations keep
their form (\ref{maxwell}), (\ref{maxtensor}), or
(\ref{meaningGR})$_1$ with (\ref{meaningGR})$_2$, respectively. There
is simply no place for a Poincar\'e covariant formulation of Maxwell's
equations \`a la (\ref{meaningSR}). A less extreme case is the Global
Positioning System. Nevertheless, also the GPS rules out Poincar\'e
covariant electrodynamics, see Ashby \cite{Ashby}.

The generally covariant and metric-free form of electrodynamics,
together with setting up an appropriate analysis of the physical
dimensions involved in mechanics and electrodynamics, has been worked
out by Kottler \cite{Kottler}, Cartan \cite{Cartan}, van Dantzig
\cite{vanDantzig}, Schouten \& Dorgelo \cite{SchoutenPhysicists},
Toupin \& Truesdell \cite{Truesdell}, and Post \cite{Post}.  Post's
book contains most of the relevant information, see also Kovetz
\cite{Attay} and our book \cite{book}.

\section{Premetric electrodynamics}

We base electrodynamics on {\em electric charge conservation} (first
axiom), in differential form $dJ=0$ or, in tensor calculus,
$\partial_i\check{J}^i=0$. This law is metric-independent since it is
based on a {\it counting\/} procedure for elementary charges. Charge
conservation is a law that is valid in macro- as well as in
micro-physics.\footnote{L\"ammerzahl et al.\ \cite{Laemmerzahl} are
  studying extensions of Maxwell's equations that violate charge
  conservation. Such models can be used as test theories for
  experiments on the checking of charge conservation.} The same is
true for $J=dH$, a consequence of $dJ=0$ and the de Rham theorem. But
this is already the inhomogeneous Maxwell equation
(\ref{maxwell})$_1$. The excitation $H$ features as a potential of the
current $J$, that is, the inhomogeneous Maxwell equation is a
consequence of charge conservation (and not the other way round, as it
is often argued in textbooks).

The axiom of charge conservation, a metric-independent law, explains
why the inhomogeneous Maxwell equation $dH=J$ is likewise metric-free.
And, not to forget, the excitation $H$ is valid on the same level as
the current $J=(j,\rho)$. The excitation is as microscopic a field as
is $J$ (and $E$ and $B$). Since charge conservation is valid at any
level of resolution, the same is true for the existence of the
excitation $H$.  To conceive the excitation $H=({\cal H},{\cal D})$ as
only a macroscopic field, is simply neglecting the experimental
underpinning of the law of the conservation of electric charge.

We decompose the 4-dimensional excitation $H$ into two pieces: one
along the 1-dimensional proto-time $\sigma$ and another one embedded
in 3-dimensional space. We find (see \cite{book}):
\begin{equation}\label{cutexcitation}
  H=-{\cal H}\wedge d\sigma + {\cal D}.
\end{equation}
Then the 3-dimensional inhomogeneous Maxwell equations read
\begin{equation}\label{dH}
   \underline{d}\,{\cal D}=\rho\,,\qquad
   \underline{d}\,{\cal H}-\dot{{\cal D}}=j\,,
\end{equation}
i.e., we recover the Coulomb-Gauss and the Oersted-Amp\`ere-Maxwell laws.
The underline denotes the 3-dimensional exterior derivative and the
dot differentiation with respect to the proto-time $\sigma$.

With charge conservation alone, we arrived at the inhomogeneous
Maxwell equations (\ref{dH}). Now we need some more input for deriving
the homogeneous Maxwell equations. The force on a charge density is
encoded into the axiom of the {\em Lorentz force density\/} (second
axiom)
\begin{equation}\label{Lorentz}
  f_\alpha =(e_\alpha\rfloor F)\wedge J\,.
\end{equation}
Here $e_\alpha$ is an arbitrary (or anholonomic) frame or tetrad, a
basis of the tangent space, with $\alpha=0,1,2,3$, and $\rfloor$
denotes the interior product (contraction). In tensor language, we can
express (\ref{Lorentz}) as $\check{f}_i=F_{ik} \,\check{J}^k$. The
axiom (\ref{Lorentz}) should be read as an operational procedure for
defining the electromagnetic field strength 2-form $F=F_{ik}\,dx^i
\wedge dx^k/2$ in terms of the force density $f_\alpha$, known {}from
mechanics, and the current density $J$, known {}from charge
conservation.  The 1+3 decomposition of $F$ reads
\begin{equation}\label{Fdecomp}
  F=E\wedge d\sigma+B\,,
\end{equation}
with the electric and the magnetic field strengths $E$ and $B$,
respectively, see also Figure 1.  

{\em Magnetic flux conservation\/} is our third axiom.  In its local
form it reads $dF=0$. This is the homogeneous Maxwell equation
(\ref{maxwell})$_2$. Split into 1+3, we find
\begin{equation}\label{dB}
\underline{d}\,E+\dot{B}=0\,,\qquad \underline{d}\,B=0\,,
\end{equation}
i.e., Faraday's induction law\footnote{The Lenz rule and the reason
  for the relative sign difference between the time derivatives in
  (\ref{dH})$_2$ and (\ref{dB})$_1$ are discussed in \cite{ItinLenz}.}
and the sourcelessness of $B$. The laws (\ref{dB}) are also
metric-free since at least in certain situations, namely in
superconductors of type II, magnetic flux lines can be {\it counted}.

\begin{figure}
\epsfxsize=\hsize 
\epsfbox{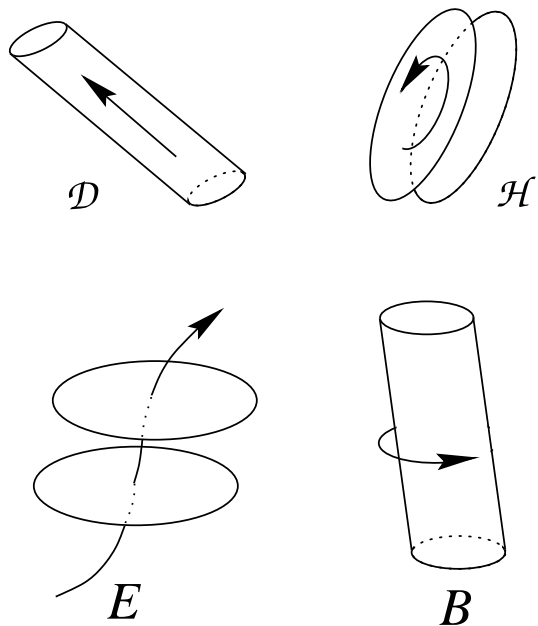}
\caption[Faraday--Schouten pictograms of the electromagnetic field $(H,F)$
in 3-dimensional space, see \cite{book}. Depicted are electric
excitation $\cal D$, magnetic excitation $\cal H$ and electric field
strength $E$, magnetic field strength $B$. These four 3-dimensional
electromagnetic fields are described by {\it four different\/}
geometrical objects that are defined by means of the group of general
coordinate transformations. The images of 1-forms are represented by
two neighboring planes. The nearer the planes, the stronger the 1-form
is.  The 2-forms are pictured as flux tubes. The thinner the tubes,
the stronger the flow. The difference between a twisted and an
untwisted form accounts for the two different types of 1- and 2-forms,
respectively.]  {\label{schouten}Faraday--Schouten pictograms of the
  electromagnetic field $(H,F)$ in 3-dimensional space, see
  \cite{book}.  Depicted are electric excitation $\cal D$, magnetic
  excitation $\cal H$ and electric field strength $E$, magnetic field
  strength $B$.  These four 3-dimensional electromagnetic fields are
  described by {\em four different\/} geometrical objects that are
  defined by means of the group of general coordinate transformations.
  The images of 1-forms are represented by two neighboring planes. The
  nearer the planes, the stronger the 1-form is.  The 2-forms are
  pictured as flux tubes.  The thinner the tubes, the stronger the
  flow. The difference between a twisted and an untwisted form
  accounts for the two different types of 1- and 2-forms,
  respectively.}
\end{figure}

All our considerations in this section are generally covariant and
metric-free and connection-free. They are valid in flat Minkowskian
and in curved Riemannian spacetime, that is, in SR and in GR. Even if
spacetime carried torsion and/or nonmetricity, see
\cite{Preuss1,Rubilar,Itin,Solanki,Preuss2}, there would be no need to
reformulate Maxwell's equations. Therefore Maxwell's equations in the
form of (\ref{maxwell}), of (\ref{maxtensor}) or of (\ref{dH},\ref{dB})
represent the optimal formulation of the fundamental laws of classical
electrodynamics. Before we turn to the relation between $H$ and $F$,
we will have a look at dimensional analysis.

\section{Dimensional analysis and the generally covariant 4D scalars 
  of charge and action}

A physical quantity is qualitatively characterized by a certain
dimension, as, e.g., by the dimension of {\it length, mass, action,
  momentum,} or {\it electric charge.} The dimension is a reminder of
the measuring procedure for this quantity. Length can be measured by a
micrometer, momentum by a collision process, an electric current by an
ammeter, etc. Accordingly, the dimension of a quantity may be
understood as shorthand notation symbolizing the measurement procedure
for this quantity. This, in our opinion, down-to-earth approach may be
contrasted with Veneziano's view \cite{dov} who wrote: ``... it looks
unnecessary (and even ``silly" according to the present understanding
of physical phenomena) to introduce a separate unit for temperature,
for electric current and resistance, etc..."

A certain domain of physics has a small number of {\it base\/}
dimensions and other dimensions {\it derived\/} therefrom as products
or quotients with powers in terms of the base dimensions.
Acceleration, e.g., can be expressed as $\ell/t^2$, where ``dimension
of length'' is abbreviated as $\ell$ and ``dimension of time'' as $t$.
In mechanics we have three base dimensions, length $\ell$, time $t$,
and action $h$.  Usually, instead of the action, the mass is taken as
a base dimension.  However, {}from a principal point of view (in
nature, there exists a universal constant with the dimension of an
action) and {}from a practical point of view (an action can be more
precisely measured than a mass), the dimension of action $h$ as base
dimension is to be preferred. We follow here the dimensional analysis
of Schouten and Dorgelo, see \cite{SchoutenPhysicists}, and, in
particular, of Post \cite{Post}. In electrodynamics, besides the base
dimensions of mechanics, namely $\ell$, $t$, and $h$, we introduce the
base dimension of {\it charge\/} $q$, see Sommerfeld \cite{Sommerfeld}:
\begin{equation}\label{basedim}
  \text{base dimensions}\;\rightarrow\;(\ell,\,t,\,h,\,q)\,.
\end{equation}
Action and charge are (4-dimensional) scalars in spacetime.  Thus the
(anholonomic) components of the electric field, for example, have the
dimension $[E_a]=h/(q\,t \,\ell)$, see \cite{book}.

Vectors, forms, and other physical quantities with more than one
component require special consideration. Already the electric field,
just discussed, is such a case. We will come back to it. Because here
we are mainly interested in electrodynamics, we concentrate on
differential forms. In the framework of exterior calculus, the notions
of absolute and relative dimensions \cite{SchoutenPhysicists}, which
we are going to introduce, become almost trivial. Whenever we have a
$p$-form $\psi$, we may take an arbitrary $p$-dimensional submanifold
$S$ of spacetime and calculate the integral $\int_S\psi$.  This is a
quantity that does not depend on the choice of the local coordinates
nor of the frames. The dimension of that quantity
$\left[\int_S\psi\right]$ is called the {\it absolute} dimension of
$\psi$. The {\it relative\/} (or physical) dimension is the dimension
$[\psi_{\a_1\a_2\dots \a_p}]$ of its components with respect to a
coframe $\vta^\a$. Here $[\vta^0]=t\,,\;[\vta^a]=\ell$, with
$a=1,2,3$.

Each physical quantity $Q$ consists of its magnitude $\{Q\}$ and its
dimension $[Q]$ such that $Q=\{Q\}\times[Q]$, see also Massey
\cite{Massey}, Sedov \cite{Sedov}, Sec.7.7, and Wallot \cite{Wallot}.
An equation that is built up {}from physical quantities --- this is
particularly valid for a law of nature --- is called a {\it quantity
  equation.} Its left-hand-side is represented by a magnitude and a
dimension; the same is true for its right-hand-side. Such an equation
is required to be dimensionally consistent. We would like to stress
that these dimensional considerations are valid independent of the
units that we may eventually want to choose. Physicist who are
comfortable with ``physical'' equations consisting only of sheer
numbers, have to arrange beforehand a table in which all quantities
are enlisted and the allowed units specified in terms of which the
numbers are to be determined. This is a possibility, but it tends to
hide the dimensions of the quantities involved and the ``number
equations'' becomes opaque and unilluminating {}from a dimensional
point of view.  Quantity equations are valid for arbitrarily chosen
units and no prearrangements are necessary.

Let us come back to premetric electrodynamics.  To begin with, we
consider the electric current 3-form $J$. Its integral over an
arbitrary 3-dimensional domain describes the total charge contained
therein. It can be determined by counting the particles carrying
an elementary charge. The latter can be taken as a unit of charge, in
the sense of the theory of dimensions.

Thus, the {\em absolute\/} dimension of $J$ is that of a charge:
$[J]=q$.  Since the exterior derivative is dimensionless, $[d]=1$, and
since the electromagnetic excitation is given by the inhomogeneous
Maxwell equation $dH = J$, we conclude that the absolute dimension of
the excitation $[H] = q$. Then, by means of (\ref{cutexcitation}), the
absolute dimensions of the electric and magnetic excitations turn out
to be $[{\cal D}]=q$ and $[{\cal H}]=q/t$.  The {\em relative\/}
dimensions are those of their frame components, $[{\cal
  H}_a]=q/(t\,\ell)$ and $[{\cal D}_{ab}]=q/\ell^2$, with the spatial
indices $a,b,=1,2,3$.  These are the ``physical'' dimensions known to
physicists and engineers.

Since we denoted the physical dimension of an action by ${h}$,
equation (\ref{Lorentz}) shows that the absolute dimension of the
electromagnetic field strength 2-form $F$ is $[F]=h/q=\phi$, that is,
action/charge or magnetic flux $\phi$, see \cite{book}. Then by
(\ref{Fdecomp}), the absolute dimensions of the electric and magnetic
fields $E$ and $B$ are $[E]=\phi/t$ and $[B]=\phi$, respectively, and
the relative dimensions $[E_a] =\phi/(t\,\ell)$ and
$[B_{ab}]=\phi/\ell^2$. Again, these are the every-day dimensions
known to physicists and engineers alike, see Table 1. The (passive)
quantities related to force carry a magnetic flux in their
dimensions, the (active) ones related to charge the dimension of
charge. This is nontrivial information that is buried if one
suppresses dimensions.
\bigskip

\begin{table}[htbp]
\begin{center}
  Table 1. The physical dimensions of the electromagnetic field. The
  dimensions are abbreviated as follows: $\phi\rightarrow$ magnetic
  flux, $q\rightarrow$ electric charge, $t\rightarrow$ time, and
  $\ell\rightarrow$ length.\\ 
\bigskip

{\begin{tabular}{|c||c|} \hline absolute& relative\\ \hline\hline
      {\begin{tabular}{c|c} &\\ $[E]=\frac{\phi}{t}$ &$[B]=\phi$\\ &\\ 
          \hline &\\ $[{\cal H}]=\frac{q}{t}$&$[{\cal D}]=q$\\ &
\end{tabular}} & {\begin{tabular}{c|c}&\\
$[E_a]=\frac{\phi}{t\,\ell}$&$[B_{ab}]=\frac{\phi}{\ell^2}$\\ &\\ \hline  &\\
 $[{\cal H}_a]=\frac{q}{t\,\ell}$&$[{\cal D}_{ab}]=\frac{q}{\ell^2}$\\ &
\end{tabular}} \\ \hline
\end{tabular}}
\end{center}
\end{table}

The charge defined by the integral $\int J$ is an {\it invariant}
quantity under general coordinate and frame transformations on the
spacetime manifold. Similarly, the integral of the Lagrangian 4-form
$V$ of the (electromagnetic or matter) field over a 4-dimensional
domain $\int V$ is an {\it invariant} quantity with the dimension of
action $h$. Thus, the dimensions of charge and action $q, h$ are
distinguished {}from other physical dimensions, such as mass, length,
and time.
Accordingly, {\em electric charge\/} and {\em action} are 4D
invariants or scalars.  Consequently,\footnote{Post \cite{Posthbar}
  has an argument that {\em action\/} is a {\it pseudo\/}scalar. If
  the same were true for {\em charge,} then this would be consistent
  with the twisted nature of $H$ and the untwisted one of $F$, since
  $[F]=[\mbox{action/charge}]=h/q$. Such a prescription could exclude
  noninteger numbers $n_1$ and $n_2$ in (\ref{scalars}).} a premetric
physical quantity with dimensions
\begin{equation}\label{scalars}
  q^{n_1}\,h^{n_2}\,=\,\mbox{4D scalar}
\end{equation}
is a very natural structure. Strictly, by our arguments $n_1$ and $n_2$
are not required to be integers. However, examples for such dimensionful
4-scalars are
\begin{equation}\label{examplesSC}
  q\rightarrow\mbox{electric charge}\,,\quad \frac{h}{q}\rightarrow
  \mbox{magnetic flux}\,, \quad
  \frac{h}{q^2}\rightarrow\mbox{electric resistance}\dots\,.
\end{equation}
And there we only observe $n_1,n_2=0,\pm1,\pm2,\dots$ It is remarkable
that modern metrology \cite{Flowers,metro1,CODATA,metro2}, making use
of the Josephson and of the quantum Hall effects, provide highly
precise measurements of the {\it Josephson\/}
\cite{Josephson,Andreone} and the {\it von Klitzing\/}
  \cite{Klitzing,Jeckelmann} constants, respectively:
\begin{equation}
  K_{\rm J} = {\frac {2e} {\rm h}},\qquad R_{\rm K} = {\frac {\rm h}
    {e^2}}.\label{KR}
\end{equation}
Here $e$ denotes the elementary charge and {\rm h} Planck's constant.
These two quantities are both of type (\ref{scalars}). The premetric
nature of $R_{\rm K}$ has helped us to predict \cite{QHEgrav} that the
quantum Hall effect cannot couple to the gravitational field.

So far, our dimensional analysis did not make use of the metric.
Moreover, it is generally covariant and as such valid in particular in
GR and SR. And, on top of that, our considerations do not depend on
any particular choice of the system of physical units. Whatever your
favorite system of units may be, our results will apply to it. In
short: Our dimensional analysis so far is {\it premetric, generally
covariant, and valid for any system of units.}

\section{Vacuum resistance $\Omega_0$ and the speed of light $c$ 
  as properties of spacetime}

The spacetime metric naturally enters electromagnetic theory in the
constitutive relations between the excitation and the field strength,
$H = H(F)$, see (\ref{meaningGR})$_3$.  A local and linear spacetime
relation is the natural first choice:
\begin{equation}\label{locallinear}
H_{ij} = {\frac 12}\,\kappa_{ij}{}^{kl}\,F_{kl}\,.
\end{equation}
The $6\times 6$ functions $\kappa_{ij}{}^{kl}(x)$ form the twisted
constitutive tensor of spacetime. 

Up to quadratic order, we can construct two 4-dimensional {\em
  invariants\/} {}from this object:
\begin{equation}
\alpha := {\frac 1{12}}\,\kappa_{ij}{}^{ij},\qquad \lambda^2 := 
-\,{\frac 1{4!}}\,\kappa_{ij}{}^{kl}\kappa_{kl}{}^{ij}.\label{allad}
\end{equation}
The signs and the numeric factors are conventional. It is important
that these objects are both scalar quantities with definite absolute
dimensions.  Recalling the dimensions of the excitation and the field
strength, we find
\begin{equation}
  [\alpha] = [\lambda] = {\frac {q}{\phi}} = {\frac
    {q^2}{h}}\;\rightarrow\;\mbox{1/electric resistance}\,.\label{alla}
\end{equation}
This fact is of fundamental importance since it is {\it independent}
of any choice of the local frames or coordinates.\footnote{Recently we
  developed a method for deriving the light cone from a number of
  requirements imposed on the local and linear spacetime relation
  (\ref{locallinear}), see \cite{HOPeres} and references given there.}

The standard Maxwell-Lorentz electrodynamics arises when we assume
additionally {\em isotropy\/} relations between the (extensive, or
additive) quantities $\cal D$ and $\cal H$ and the field intensities
$E$ and $B$, respectively. The Maxwell-Lorentz spacetime relation is
specified by
\begin{equation}
  {\cal D} = \varepsilon_0\;{}^{\underline\ast} E\qquad{\rm and}\qquad
  {\cal H} = {\frac 1 {\mu_0}}\;{}^{\underline\ast}
  B\,.\label{linearlaw}
\end{equation}
The 3-dimensional Hodge star operator ${\underline\ast}$ is needed here for 
the mapping of a 1-form into a 2-form and vice versa or, more generally, a 
$p$-form into a $(3-p)$-form. 

The proportionality coefficients $\varepsilon_0$ and $\mu_0$, together
with the Hodge star ${\underline\ast}\,$, encode all essential
information about the electric and magnetic properties of spacetime.
This is where the metric of spacetime enters the theory of
electromagnetism. Indeed, defining
\begin{equation}
c := {\frac 1{\sqrt{\varepsilon_0\mu_0}}}\qquad {\rm and}\qquad 
\Omega_0 := \sqrt{\frac{\mu_0}{\varepsilon_0}}\,,\label{cOmega}
\end{equation}
we can rewrite (\ref{linearlaw}) in matrix form
\begin{equation}
\left(\begin{array}{c}{\cal H}\\ {\cal D}\end{array}\right) = 
\lambda_0\left(\begin{array}{cc} 0 & c\;^{\underline\ast} \\ 
c^{-1\;\underline\ast} & 0\end{array}\right) 
\left(\begin{array}{c} E\\ B\end{array}\right),\label{ML}
\end{equation}
with $\lambda_0 := 1/\Omega_0$. Comparing with (\ref{locallinear}), we 
find that the Maxwell-Lorentz spacetime relation is achieved by means 
of the constitutive tensor density
\begin{equation}\label{kappametric}
  \kappa_{ij}{}^{kl}=\lambda_0\,\hat{\epsilon}_{ijmn}\,
  \sqrt{-g}\,g^{mk}g^{nl}\,,
\end{equation}
where $\hat{\epsilon}_{ijmn}$ is the Levi-Civita symbol and $g_{ij}$
the spacetime metric; in Cartesian coordinates, the latter reads
$g_{ij}={\rm diag} (c^2, -1, -1, -1)$. As we can verify, the
invariants (\ref{allad}) are now $\alpha =0$ and $\lambda =
\lambda_0=1/\Omega_0$. An equivalent 4-dimensional form of
(\ref{linearlaw}) or (\ref{ML}) can be written, with the help of the
(4-dimensional) Hodge star $\star$ of the metric $g_{ij}$, as
\begin{equation}
H = \lambda_0{}\,^\star\!F\,.\label{spacetimerel}
\end{equation}
We may consider this law as constitutive relation for spacetime
itself, with the ``moduli'' $c$ and $\Omega_0$ ($=1/\lambda_0$).

The operator ${\underline\ast}$ in (\ref{linearlaw}) has the dimension
of a length $\ell$ or its reciprocal $1/\ell$, respectively.
Recalling the {\it dimensions\/} of
the excitations and the field strengths, we find the dimensions of the 
electric constant $\varepsilon_0$ and the magnetic constant $\mu_0$ as
\begin{equation}
[\varepsilon_0]={\frac{q\,t}{\phi\,\ell}}\qquad 
{\rm and}\qquad[\mu_0]={\frac{\phi\,t}{q\,\ell}}\,,\label{epsmu0}
\end{equation}
respectively. They are also called vacuum permittivity and vacuum
permeability, see the new Codata report \cite{CODATA}. 
Dimensionwise, it is clearly visible that the two quantities introduced
in (\ref{cOmega}) are 
\begin{equation}
[c] = \ell/t\qquad {\rm and}\qquad [\Omega_0] = \phi/q = h/q^2\,. 
\end{equation}
Obviously, the velocity $c$ and the resistance $\Omega_0$ are
constants of nature, the velocity of light $c$ being a universal one,
whereas $\Omega_0$, the characteristic impedance (or wave resistance)
of the vacuum, 
seemingly refers only to the electromagnetic properties of spacetime.
Note that $\lambda_0$ plays the role of the coupling constant of the
electromagnetic field which enters as a factor the free field Maxwell
Lagrangian
\begin{equation}\label{Lagr}
  V = -\,{\frac 12} \lambda_0\,F\wedge{}^\star\!F\,.
\end{equation}

The Maxwell equations (\ref{maxwell}), together with the
Maxwell-Lorentz spacetime relation (\ref{spacetimerel}) [or, in $1+3$
form, (\ref{dH}) and (\ref{dB}), together with (\ref{linearlaw})],
constitute the foundations of classical electrodynamics:
\begin{equation}
d\;^\star \!F=\Omega_0\,J\,,\qquad dF=0\,.\label{maxwell1}
\end{equation}
These laws, in the classical domain, are assumed to be of {\it
  universal validity}. Only if vacuum polarization effects of quantum
electrodynamics are taken care of or if hypothetical nonlocal terms
emerge due to huge accelerations, the spacetime relation $H = H(F)$
can pick up corrections yielding a {\it nonlinear\/} law
(Heisenberg-Euler electrodynamics \cite{HeisenbergEuler}) or a {\it
  nonlocal\/} law (Volterra-Mashhoon electrodynamics, see
\cite{Bahram0,Bahram1,Muench}), respectively, see also \cite{book}.
In this sense, the Maxwell equations (\ref{maxwell}) are ``more
universal'' than the Maxwell-Lorentz spacetime relation
(\ref{spacetimerel}). The latter is not completely untouchable.

Let us underline that everything derived in this section is
independent of any choice of the system of units. We are relying on
general dimensional analysis alone.

\section{Time dependence of fundamental constants}

The possibility of time and space variations of the fundamental
constants is discussed in the literature both from an experimental and
a theoretical point of view, see \cite{Karshenboim,Honnef,Uzan}, for
example.  Of particular interest are certain indications that the fine
structure constant may slowly change on a cosmological time scale. It
is thus important to know whether this fact can be related to a
possible variation of the physical constants: ${\rm h}$, $e$, $c$, or none
of these?

Maxwell's equations follow from charge and flux conservation. Any
charge is proportional to the elementary charge $e$, any flux is
proportional to the elementary flux ${\rm h}/e^2$. Consequently, if $e$
and ${\rm h}$ keep their values constant (independent of time, e.g.), then 
the quantities proportional to them or any power of them, namely $e^{n_1}
{\rm h}^{n_2}$, with $n_1$ and $n_2$ as integer numbers, are also conserved.
Therefore the time independence of $e$ and ${\rm h}$ are the raison
d'etre for the Maxwell equations. Or the other way round: If we want
to uphold the Maxwell equations, then we have to demand $e=\mbox{const}$ 
and ${\rm h}=\mbox{const}$. 

As we already mentioned above, charge $q$ and action $h$ are 4D
scalars. Thus no time dependence is allowed provided the premetric
Maxwell equations are assumed to be valid. This, however, is not true
for $c$, and that was the reason why Peres \cite{Peres} related the
experimental evidence of the variability of the fine structure
``constant'' to the change of the speed of light $c=c(t,x^a)$.

However, there is a different possibility. In order to realize this, let
us have a closer look at the definition of the fine structure constant:
\begin{equation}\label{fine}
\alpha_{\rm f} = {\frac {e^2}{2\varepsilon_0\,c\,{\rm h}}} = 
{\frac {e^2}{2\,{\rm h}\,\lambda_0}} = {\frac {\Omega_0}{2R_{\rm K}}}.
\end{equation}
As we see, the fine structure constant is explicitly given in terms of
the ratio of two resistances --- vacuum impedance $\Omega_0$ and von
Klitzing constant $R_{\rm K}$ (the quantum Hall resistance). Note that
the speed of light $c$ {\it disappeared completely!} Its ``presence"
in the first equality is in fact misleading and the proper
understanding is suggested only in the second equality where
$\lambda_0$ shows up instead, together with $e$ and ${\rm h}$.


In other words, the formula (\ref{fine}) demonstrates that of the two
fundamental constants of electrodynamics, which appear naturally in
the Maxwell-Lorentz electrodynamics (see the previous section), it is
the vacuum impedance which enters the fine structure constant and {\it
  not} the speed of light.

Recall now again the argument \cite{Peres} that $e$ and ${\rm h}$,
being 4D scalars, should not change in time and space provided one
wants to uphold the validity of the Maxwell equations (\ref{maxwell}).
Then the variation of the fine structure constant $\alpha_{\rm f} =
\alpha_{\rm f}(t)$ forces us to conclude that $\lambda_0 =
\lambda_0(t)$. An inspection of the Maxwell Lagrangian (\ref{Lagr})
then shows that $\lambda_0$ becomes a dynamical {\it dilaton\/} field.
Such models were studied by Bekenstein \cite{Bekenstein}, although
with a different physical interpretation (of a variable $e$).  In the
axiomatic pre-metric approach to electrodynamics, we have two
(pseudo)scalar parts of the spacetime relation, which are independent
of the spacetime metric: these are the dilaton and the axion. The
variability of the fine structure constant thus may be explained by
the presence of the {\it dilaton\/} field in the Maxwell-Lorentz
spacetime relation.

\section{Choice of units: here SI}

We want to stress that the electromagnetic ``moduli'' of spacetime
(of ``vacuum''), namely the speed of light $c$ and the vacuum
resistance $\Omega_0$, can be identified by a generally covariant
dimensional analysis, which is valid in GR and SR likewise. By the
same token, the electric and the magnetic constants $\varepsilon_0$
and $\mu_0$, see (\ref{cOmega}), emerge in a consistent
approach to electrodynamics willy nilly and are, in particular, not
attached to any specific system of physical units. The absolute or
relative dimensions of a physical quantity are inborn and cannot be
chosen freely. As long as physics remains an experimental science, the
notion of the dimension of a physical quantity will be with us.

In the end, however, we want to pick a convenient system of units. We
will choose the SI-system. It is well-known from the literature how
one can go over to other system of units, see, e.g., Massey
\cite{Massey}, Post \cite{Post}, or Sommerfeld \cite{Sommerfeld}. The
base dimensions of the SI-system for mechanics and electrodynamics are
$(\ell, M, t, q/t)$, with $M$ as dimension of mass. The units are
meter, kilogram, second, and ampere, respectively (therefore
originally called the MKSA-system).

However, if we choose as base dimensions (\ref{basedim}), we find the
following SI-units:
\begin{equation}\label{lthq}
(\ell,\,t,\,h,\,q)\;\rightarrow\;(m,\, s,\, W\!b\!\times\!C,\, C)\,.
\end{equation}
Thus, instead of the base units kilogram and ampere, we choose
joule$\,\times\,$second (or weber$\,\times\,$coulomb) and coulomb.
Accordingly, coulomb and weber, that is, electric charge and magnetic
flux take center stage, as it should be in a domain of physics where
charge and flux conservation are at its basis. Numerically, in the
SI-system, one puts (for historical reasons)
\begin{equation}
\mu_0=4\pi\times10^{-7}\;\frac{W\!b\,s}{C\,m}\qquad
{\rm (magnetic\>constant)}\,.\label{mu0}
\end{equation} 
Then measurements \`a la Weber-Kohlrausch \cite{WK}, see Sec.5.3.1 of
\cite{Raith}, yield
\begin{equation}
\varepsilon_0=8.854\;188\times 10^{-12}\;\frac{C\,s}{W\!b\,m}
\qquad{\rm (electric\>constant)}\,.\label{eps0}
\end{equation}

\section{Comments on a recent paper of Okun \cite{Okun}}

First of all we should say that Okun, at least in the realm of
elementary particle physics, expresses the majority view as it is
given, e.g., in the textbooks of Feynman \cite{Feynman}, Jackson
\cite{Jackson}, or Landau-Lifshitz \cite{Landau}, see also the
discussions of Chambers \cite{Chambers} and Roche \cite{Roche}. We
believe, however, see Secs.IV, V, and VI above and Ref.\cite{book},
that the premetric approach embodies a number of decisive advantages
that leads to a better understanding of the structure of classical
electrodynamics as compared to the conventional approach.

Our three main points against Okun's Sec.10 on electrodynamics are the
following:

\subsection{General covariance and Maxwell's equations}

In Okun's discussion on electrodynamics in his Sec.10, GR is never
mentioned even though the title of this section reads
``Electromagnetism and Relativity.''\footnote{In Okun's Sec.11,
  ``Concluding remarks,'' the Global Positioning System is cited and
  its relation to GR. However, it is apparently not conceived as a
  problem related to ({\em non-\/}Poincar\'e covariant)
  electrodynamics.} Okun exclusively uses SR for the discussion of
Maxwell's equations.  This was appropriate in 1908.

It is true, as Okun mentions, that Maxwell related his fields $E,{\cal
  D},{\cal H},B$ to the aether.  Already in 1920, Einstein
\cite{EinsteinLeyden} pointed out in his Leiden lecture that the
gravitational potential $g_{ij}$ is some kind of new aether
reminiscent of the old aether of the 19th century.  However, one must
not attribute a velocity to the new aether at each point of spacetime.
To quote Einstein's summary (our translation): ``If we follow general
relativity theory, space is endowed with physical qualities; thus, in
this sense, there exists an aether. According to general relativity
theory, space without aether is unthinkable; for in such a space there
would be not only no propagation of light but also no possibility for
measuring rods and clocks to exist, hence also no spatial and temporal
intervals in the sense of physics. However, this aether must not be
thought of being endowed with the characteristic properties of
ponderable matter, namely to consist of parts that can be tracked
through time; the concept of motion must not be applied to it.''

We take Einstein's point of view. One should also note that Truesdell
\& Toupin \cite{Truesdell} and Kovetz \cite{Attay} call $H=H(F)$
aether relation (we chose the historically more neutral expression
``spacetime relation'').

If one knows the spacetime relation (\ref{locallinear}) with
(\ref{kappametric}) or, more compactly, in exterior calculus,
(\ref{spacetimerel}), then one can eliminate $H$ {}from the Maxwell
equations and ends up with (\ref{maxwell1}). Then $F=(E,B)$ is left
over and the gravitational field is encoded into the Hodge star
$^\star$. However, with equal right one could eliminate $F$. The Hodge
dual of (\ref{spacetimerel}) reads
$F=-\Omega_0\,^\star H\,.$
Thus,
\begin{equation}
dH=J\,,\qquad d\,^\star H=0\,.\label{maxwell2}
\end {equation}
Now $H=({\cal H},{\cal D})$ is left over, like in Lorentz's electron
theory of 1904 \cite{Lorentz1904}. However, {\it by no covariant
  means\/} one is ever led to Okun's choice of the pair
$\mathbf{E},\mathbf{H}$, unless one enforces some strange units in
order to accommodate $\mathbf{E}$, $\mathbf{H}$ as fundamental
variables in electrodynamics. With $\mathbf{E}$ and $\mathbf{H}$ one
cannot build up a generally covariant quantity. One is caught in flat
spacetime in Cartesian coordinates with strange units.

The book of Sommerfeld \cite{Sommerfeld} that, according to Okun, is
``prerelativistic'' even contains a discussion of GR and the
Schwarzschild solution, in contrast to Okun's {\it pre}-general
relativistic discussion of electrodynamics in his Sec.10. Okun refers
to the ``spirit of special relativity'' and what this spirit tells him
in regard to $\mathbf{E},\mathbf{B}, \mathbf{H},\mathbf{D}$, namely to
put $\mathbf{D}=\mathbf{E}$ and $\mathbf{H}=\mathbf{B}$, but we look
in vain for a corresponding enlightenment by the ``spirit of general
relativity.''

\subsection{Theory of dimensions}

Okun doesn't present a theory of dimensions as, e.g., Schouten \&
Dorgelo \cite{SchoutenPhysicists} or Post \cite{Post} do. Accordingly,
he doesn't distinguish between dimensions and units. An SI-unit can be
badly chosen. This does {\em not\/} imply, however, that the
corresponding dimension is defective. One should compare Sommerfeld
\cite{Sommerfeld} for a clear distinction between dimensions and
units.  Okun's attack against SI --- see the last but one sentence of
his abstract --- is without proper theoretical foundation.

Let us compare the dimensions of the electromagnetic field in Okun's
approach to the ones in our approach. Okun:
$[\mathbf{E}]=[\mathbf{H}]$. This is apparently meant for Cartesian
coordinates. We are not told what the dimension is. It is gratifying
to learn that the dimension of the 3D vector field $\mathbf{E}$
(electric field strength) is the same as that of the 3D vector field
$\mathbf{H}$ (magnetic excitation)! This will help our experimental
colleagues greatly. The SI-unit used by Okun for $\mathbf{H}$
is\footnote{See two lines before Okun's eq.(7).} {\em tesla}. This is
incorrect, it is $A/m$.

Incidentally, Okun is here in good company, his well-known
experimental colleague Fitch \cite{Fitch} comes up with the following:
''... any system that gives $\mathbf{E}$ and $\mathbf{B}$ different
units, when they are related through a relativistic transformation, is
on the far side of sanity."  Obviously Fitch, very much like Okun,
doesn't consider dimensions nor distinguishes an absolute {}from a
relative dimension. To repeat: The 4D field strength F has absolute
dimension $[F]= \phi$ (magnetic flux).  But in 3D, we have the
absolute dimensions $[E]= \phi/t\stackrel{\rm SI}{=}V$,
$[B]=\phi\stackrel{\rm SI}{=}Vs=W\!b $ and the relative dimensions
$[E_a]= \phi/(t\,\ell)\stackrel{\rm SI}{=}V/m$,
$[B_{ab}]=\phi/\ell^2\stackrel{\rm SI}{=}W\!b/m^2=tesla=T$, see Table
1. Hence clearly $\mathbf{E}$ and $\mathbf{B}$ do have different
dimensions and different unit and we are apparently ``on the far side
of sanity."

It should be stressed that our article is not in any sense ``pro-SI''.
It is rather aimed at a clean discussion of the theory of dimensions
that leads to the electric and the magnetic constants $\varepsilon_0$
and $\mu_0$, inter alia. A priori, this has nothing to do with any
choice of units. One shouldn't mix these two things. A lot of Okun's
anti-SI discussion is directed against an appropriate dimensional
analysis.  The real question is whether one prefers "numerical
equations" or "quantity equations". Okun opts for the first choice, we
for the second one. Specifically in SI, also the point of view of
quantity equations is adopted.

\subsection{Vacuum polarization and vacuum
  permittivity/permeability}

Vacuum polarization described by QED emerges in a special relativistic
context.  By the same token, the notions of permittivity and
permeability of the vacuum are consistent with SR.

Okun: ``Let us stress that this polarization has nothing to do with
purely classical non-unit values of $\varepsilon_0$ and $\mu_0$.''
Heisenberg and Euler \cite{HeisenbergEuler}, who ``invented'' the
subject, had a different opinion: Vacuum polarization in QED, in a
semi-classical approximation, can be understood such that Max\-well's
equations in terms of $E,{\cal D},{\cal H}, B$ remain the same, but
the relations between ${\cal D}$ and $E$ and ${\cal H}$ and $B$ get
modified, see \cite{HeisenbergEuler}, Eqs.(1) and (2). Clearly, the
spacetime relations
(\ref{linearlaw}) are perturbed by the vacuum polarization of QED,
provided a semiclassical approximation is sufficient.

\section{Conclusions}

The appropriate structure and the real beauty of classical
electrodynamics is, in our opinion, expressed in the generally
covariant equations (\ref{maxwell}) and (\ref{spacetimerel}):
$dH=J,\>dF=0$ and $H=\,^\star F/\Omega_0$.  Okun's equations
Ref.\cite{Okun}, (20,21,22) have a more restricted domain of
applicability: they are only valid in flat Minkowskian spacetime in
Cartesian coordinates. A dimensional analysis of Maxwell-Lorentz
electrodynamics yields the speed of light $c$ and the wave resistance
of the vacuum $\Omega_0$ as the two quantities characterizing the
electromagnetic properties of spacetime.

\subsubsection*{Acknowledgments} 
One of us (FWH) is most grateful to Bahram Mashhoon (Columbia,
Missouri) for many discussions on the subject of this article and to
S.G.~Karshenboim (St.Petersburg/M\"unchen) and E.~Peik (Braunschweig)
for the invitation to participate at their highly stimulating Honnef
workshop on ``Astrophysics, Clocks and Fundamental Constants''
\cite{Honnef}. We thank Lev Okun for helpful remarks on a preliminary
version of our paper. This work has been supported by the grant HE
528/20-1 of the DFG (Bonn).

\centerline{=========}

\end{document}